\documentclass[12pt]{article}
\usepackage{graphicx}
\usepackage{amsmath}
\usepackage{cite}
\usepackage[colorlinks = true,
            linkcolor = blue,
            urlcolor  = blue,
            citecolor = blue,
            anchorcolor = blue]{hyperref}

\newcommand\pubnumber{TTK-22-08, P3H-22-015}
\newcommand\pubdate{\today}

\def\institute{Institute for Theoretical Particle Physics and Cosmology\\
RWTH Aachen University, D-52056 Aachen, Germany}

\def\Title#1{\begin{center} {\Large #1 } \end{center}}
\def\Author#1{\begin{center}{ \sc #1} \end{center}}
\def\Address#1{\begin{center}{ \it #1} \end{center}}

\newcommand\pubblock{\rightline{\begin{tabular}{l} \pubnumber\\
         \pubdate  \end{tabular}}}
\newenvironment{Abstract}{\begin{quotation}  }{\end{quotation}}
\newenvironment{Presented}{\begin{quotation} \begin{center} 
             PRESENTED AT\end{center}\bigskip 
      \begin{center}\begin{large}}{\end{large}\end{center} \end{quotation}}
\def\Acknowledgements{\bigskip  \bigskip \begin{center} \begin{large}
             \bf ACKNOWLEDGEMENTS \end{large}\end{center}}

\def\beq{\begin{equation}}
\def\eeq#1{\label{#1}\end{equation}}
\def\eeqn{\end{equation}}

\def\beqa{\begin{eqnarray}}
\def\eeqa#1{\label{#1}\end{eqnarray}}
\def\eeqan{\end{eqnarray}}

\let\bar=\overbar

\def\Dslash{\not{\hbox{\kern-4pt $D$}}}
\def\dslash{\not{\hbox{\kern-2pt $\del$}}}

\def\msb{{\bar{\ssstyle M \kern -1pt S}}}

\begin{document}
\begin{titlepage}
\pubblock

\vfill
\Title{Full off-shell predictions and $b$-jet definitions for $t\bar{t}b\bar{b}$ at NLO in QCD}
\vfill
\Author{Michele Lupattelli}
\Address{\institute}
\vfill
\begin{Abstract}
In this proceeding we present predictions for the complete NLO QCD corrections to the production of $t\bar{t}b\bar{b}$ in the dilepton decay channel of the top quark at the LHC with $\sqrt{s}=13$ TeV. The calculation is performed considering all the full off-shell effects. Moreover, the various sources of uncertainty for this process are studied in detail. We then compare these predictions to experimental measurements. We additionally investigate the impact of the $b$-quark initiated subprocesses.
\end{Abstract}
\vfill
\begin{Presented}
$14^\mathrm{th}$ International Workshop on Top Quark Physics\\
(videoconference), 13--17 September, 2021
\end{Presented}
\vfill
\end{titlepage}
\def\thefootnote{\fnsymbol{footnote}}
\setcounter{footnote}{0}
\section{Introduction}

In this proceeding we present an independent computation of the complete NLO QCD corrections to the production of $t\bar{t}b\bar{b}$ in the dilepton decay channel of the top quark, namely $pp\rightarrow e^+ \nu_e \mu^- \bar{\nu}_\mu b \bar{b} b \bar{b} + X$, for LHC center of mass energy of 13 TeV. In this computation all the full off-shell effects are taken into account. This means that the heavy particles participating in the process, the top quark and the $W$ boson, are described by Breit-Wigner propagators. Moreover, all the double-, single- and non-resonant diagrams are taken into account and all the interference effects are consistently incorporated at the matrix element level.

The motivations behind this study are manifold. This process is an irreducible background to $pp\rightarrow t\bar{t}H \rightarrow t\bar{t}b\bar{b}$, which is a prime ingredient to extract information on the top-Yukawa coupling $Y_t$. However, due to the large amount of jets produced in this process, this decay channel suffers of a huge background which needs to be correctly described. The $t\bar{t}b\bar{b}$ production process is also interesting from the QCD point of view itself. Indeed, it is relevant for the precise measurement of the top-quark fiducial cross section as well as the investigation of the top-quark production and decay modelling in the presence of additional $b$-jets. Calculation with stable top quarks \cite{Bredenstein:2008zb, Bredenstein:2009aj, Bevilacqua:2009zn, Bredenstein:2010rs, Worek:2011rd, Bevilacqua:2014qfa, Buccioni:2019plc} and matched to parton shower \cite{Cascioli:2013era, Garzelli:2014aba, Bevilacqua:2017cru, Jezo:2018yaf}  are available for some time. Full off-shell calculations have been carried out only recently independently by two different groups \cite{Denner:2020orv, Bevilacqua:2021cit} and their predictions are in excellent agreement.
\section{Results}

The calculation has been carried out using the \textsc{Helac-Nlo} Monte-Carlo framework \cite{Cafarella:2007pc, Bevilacqua:2011xh}. The results have been obtained using the NNPDF3.1 PDF sets \cite{NNPDF:2017mvq}. The LO and NLO sets have been employed for the LO and NLO predictions, respectively. The 5-flavour scheme is used, meaning that the $b$ quarks are massless and present in the PDF. However, the subprocesses involving $b$ quarks in the initial state are initially neglected. Indeed, we expect these contributions to be small compared to the total fiducial cross section, both at the integrated and differential level. These contributions will be investigated in detail in the next Section. We employed both a fixed scale $m_t$ and a dynamical scale $H_T/3$, where $H_T$ is defined as follows
\begin{equation}
H_T = p_T(b_1) + p_T(b_2) + p_T(b_3) + p_T(b_4) + p_T(e^+) + p_T(\mu^-) + p_T^{miss}.
\end{equation}
The default cuts used in the calculation are
\begin{equation}
p_T(\ell) > 20 ~ \text{GeV}, ~~~ |y(\ell)| < 2.5, ~~~ p_T(b) > 25 ~ \text{GeV}, ~~~ |y(b)| < 2.5, ~~~ \Delta R(bb) > 0.4.
\end{equation}
In Table~\ref{tab:intxs} we report both the LO and NLO integrated cross sections using our two default scales. We can see that the two scales yield very similar results. Moreover, the size of the NLO QCD corrections is quite big, of the order of 89\% for the fixed scale choice and 94\% for the dynamical one. Furthermore, going from LO to NLO the theoretical uncertainties are dramatically reduced, being around 20\% at NLO. We can also notice that the main source of theoretical uncertainty comes from the scale dependence, which has been obtained using the standard 7-point scale variation. These results highlight the importance of the NLO QCD corrections for this process.
\begin{table}[]
\begin{center}
\begin{tabular}{c|cccccc}  
$\mu_0$ &  $\sigma^{\text{LO}}$ [fb] &  $\delta_{\text{scale}}$ &  $\sigma^{\text{NLO}}$ [fb] &  $\delta_{\text{scale}}$ & $\delta_{\text{PDF}}$ & $\mathcal{K}=\sigma^{\text{NLO}}/\sigma^{\text{LO}}$ \\ \hline
\rule{0pt}{3ex}
$m_t$ & $6.998$ & $^{+4.525~(65\%)}_{-2.569~(37\%)}$ & $13.24$ & $^{+2.33~(18\%)}_{-2.89~(22\%)}$ & $^{+0.19~(1\%)}_{-0.19~(1\%)}$ & $1.89$ \\
      $H_T/3$ & $6.813$ & $^{+4.338~(64\%)}_{-2.481~(36\%)}$ & $13.22$ & $^{+2.66~(20\%)}_{-2.95~(22\%)}$ & $^{+0.19~(1\%)}_{-0.19~(1\%)}$ & $1.94$ \\ \hline
\end{tabular}
\caption{\textit{Integrated fiducial cross sections for $pp\rightarrow e^+ \nu_e \mu^- \bar{\nu}_\mu b \bar{b} b \bar{b} + X$ at $\sqrt{s}=13$ TeV LHC.}}
\label{tab:intxs}
\end{center}
\end{table}

\begin{figure}[]
\centering
\includegraphics[width=0.45\textwidth]{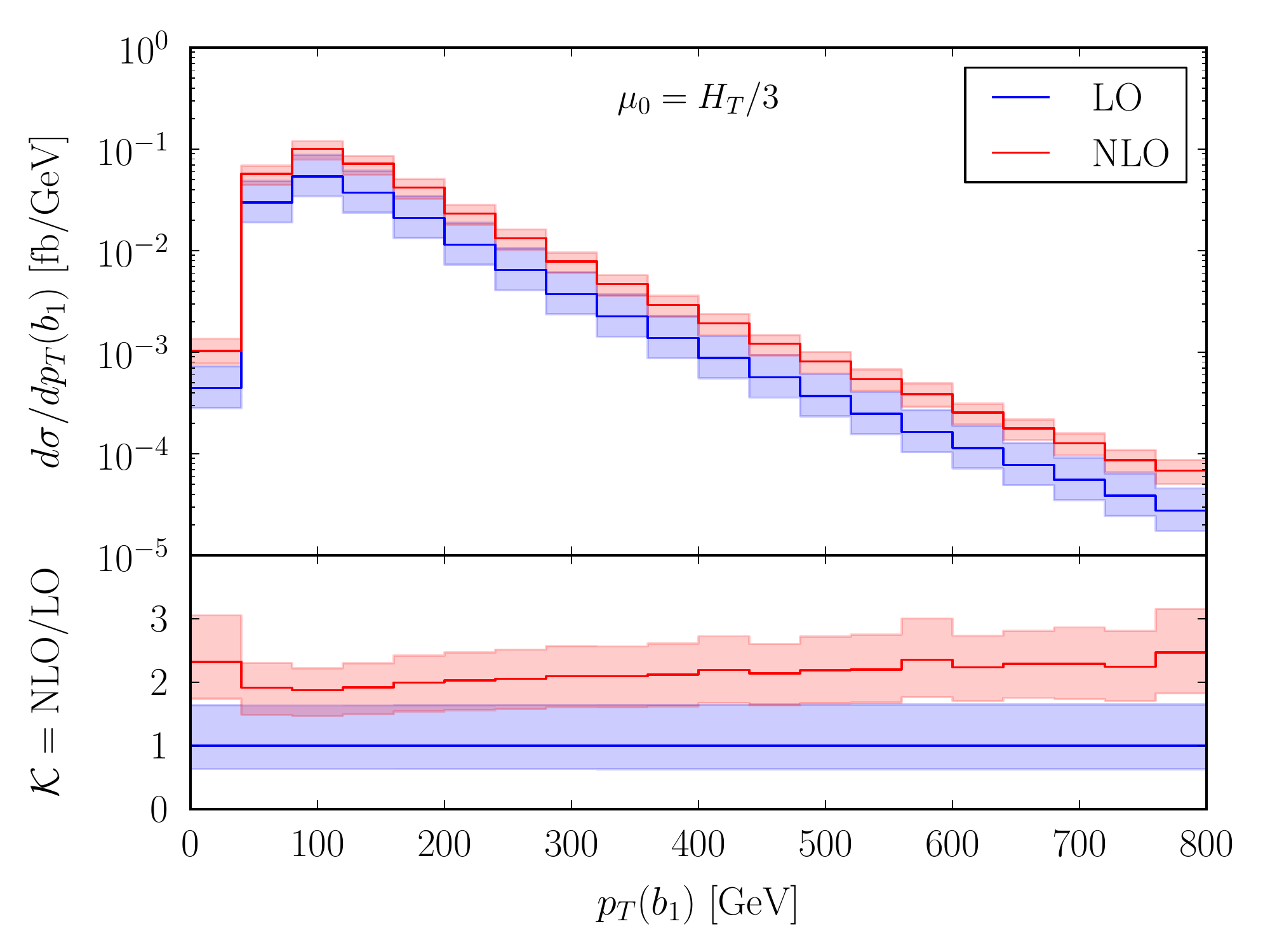}
\includegraphics[width=0.32\textwidth]{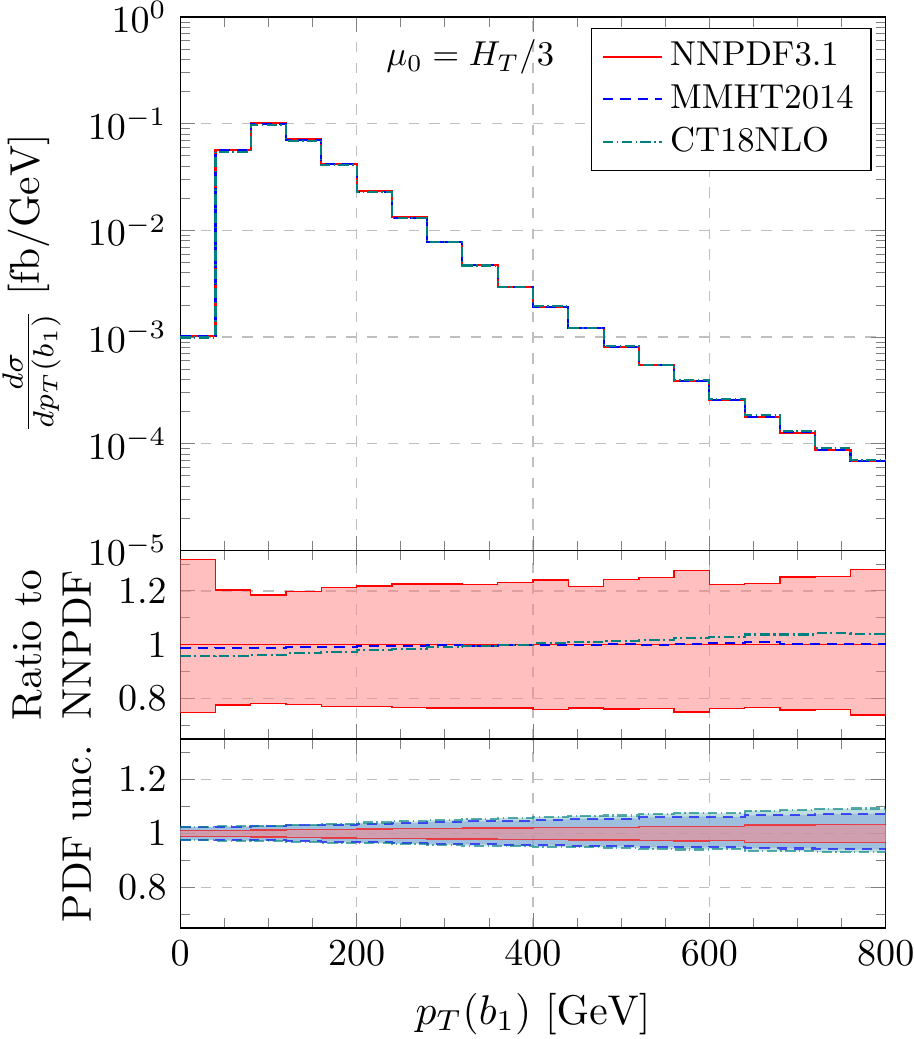}
\caption{\textit{Differential fiducial cross sections as a function of $p_T(b_1)$ for $pp\rightarrow e^+ \nu_e \mu^- \bar{\nu}_\mu b \bar{b} b \bar{b} + X$ at $\sqrt{s}=13$ TeV LHC. Figure taken from \cite{Bevilacqua:2021cit}.}}
\label{diffxs}
\end{figure}

We investigated the size of the NLO QCD corrections also at the differential level. In Figure~\ref{diffxs} we report the transverse momentum $p_T$ distribution of the hardest $b$-jet, where the $b$-jets are ordered in $p_T$. The left plot compares the LO prediction to the NLO one. The scale dependence is reported, too. We can see that also at the differential level the NLO QCD corrections are very important, and are in the range 90\%-135\%. Moreover, again, going from LO to NLO we notice a reduction of the scale dependence. The right plot reports a systematic study on the theoretical uncertainty stemming from the PDF. Various PDF sets are employed \cite{Harland-Lang:2014zoa, NNPDF:2017mvq, Hou:2019efy} and their predictions are compared to the scale dependence in the middle panel. The lower panel reports the internal PDF uncertainties. We can again conclude that the theoretical uncertainties are dominated by the scale dependence.

We then changed our setup according to Ref. \cite{ATLAS:2018fwl}, in order to compare our predictions to the experimental results obtained by the ATLAS collaboration. Some of their results have been obtained in the dilepton decay channel of the top quark. The final state signature of the leptons can either be $e^+ \mu^-$ or $e^- \mu^+$. In Table~\ref{xscomp} we provide the comparison of our prediction to these experimental results and to other theoretical predictions obtained matching fixed order calculation to parton shower algorithms, also provided in Ref. \cite{ATLAS:2018fwl}. All these results are in very good agreement within the uncertainties. The full off-shell prediction is the closest to the experimental result.
\begin{table}[]
\begin{center}
\begin{tabular}{l|c}  
Theoretical predictions & $\sigma_{e\mu+4b}$ [fb] \\ \hline
\textsc{Sherpa+OpenLoops} (4FS) & $17.2 \pm 4.2$ \\
\textsc{Powheg-Box+Pythia 8} (4FS) & $16.5$ \\
\textsc{PowHel+Pythia 8} (5FS) & $18.7$ \\
\textsc{PowHel+Pythia 8} (4FS) & $18.2$ \\ \hline
\textsc{Helac-Nlo} (5FS) & $20.0 \pm 4.3$ \\ \hline
Experimental result (\textsc{Atlas}) & $25 \pm 6.5$ \\ \hline
\end{tabular}
\caption{\textit{Comparison between several theoretical predictions and experimental results from the ATLAS collaboration.}}
\label{xscomp}
\end{center}
\end{table}

\section{$b$-quark initial states and $b$-jet tagging}
In this section we investigate the contribution of the subprocesses involving $b$ quarks in the initial state. To do so, we introduce two $b$-jet tagging schemes. The first one is the \textit{charge aware} tagging scheme. Here we assume that the jet algorithm can keep track of the charge of the $b$ jets. The recombination rules for this scheme are: $b\bar{b}\rightarrow g$, $bb\rightarrow b$, $\bar{b}\bar{b}\rightarrow \bar{b}$, $bg\rightarrow b$, $\bar{b}g\rightarrow \bar{b}$. Therefore, we require at least two $b$ jets and two $\bar{b}$ jets in the final state. The additional subprocesses we have to consider feature the following initial states: $b\bar{b}$, $bg$ and $\bar{b}g$. The second one is the \textit{charge blind} tagging scheme. As the name suggests, this scheme is blind to the charge of the $b$ jets and, therefore, the recombination rules are the following: $b\bar{b}\rightarrow g$, $bb\rightarrow g$, $\bar{b}\bar{b}\rightarrow g$, $bg\rightarrow b$, $\bar{b}g\rightarrow b$. In this case, we require at least four $b$ jets in the final state. Since the jet algorithm is blind to the charge of the $b$ quarks, we have two more possible initial states with respect to the charge aware tagging scheme: $bb$ and $\bar{b}\bar{b}$. In Table~\ref{tab:bini}, for the dynamical scale choice, we compare the integrated fiducial cross sections obtained with these two tagging schemes to the one where the $b$-quark initiated subprocesses are neglected. At LO the contribution of these subprocesses is up to 0.2\% and slightly increase to 1.2\% at NLO, due to the opening of the $bg$ and $\bar{b}g$ channels. These contributions are small compared to the theoretical uncertainties for this process and can be safely neglected.
\begin{table}[]
\begin{center}
\begin{tabular}{c|cccc}
$b$-jet tagging & $\sigma^{\rm{LO}}$ [fb] & $\sigma_i/\sigma_{\text{no }b} -1$ [\%] & $\sigma^{\rm{NLO}}$ [fb] & $\sigma_i/\sigma_{\text{no }b} -1$ [\%]  \\ \hline
      no $b$ & 6.813(3) & - & 13.22(3) & - \\
      aware & 6.822(3) & 0.1 & 13.31(3) & 0.7 \\
      blind & 6.828(3) & 0.2 & 13.38(3) & 1.2 \\ \hline
\end{tabular}
\caption{\textit{Contribution of the $b$-quark initiated subprocesses to the integrated fiducial cross section for $pp\rightarrow e^+ \nu_e \mu^- \bar{\nu}_\mu b \bar{b} b \bar{b} + X$ at $\sqrt{s}=13$ TeV LHC.}}
\label{tab:bini}
\end{center}
\end{table}
\section{Conclusion}
\begin{figure}[]
\centering
\includegraphics[width=0.4\textwidth]{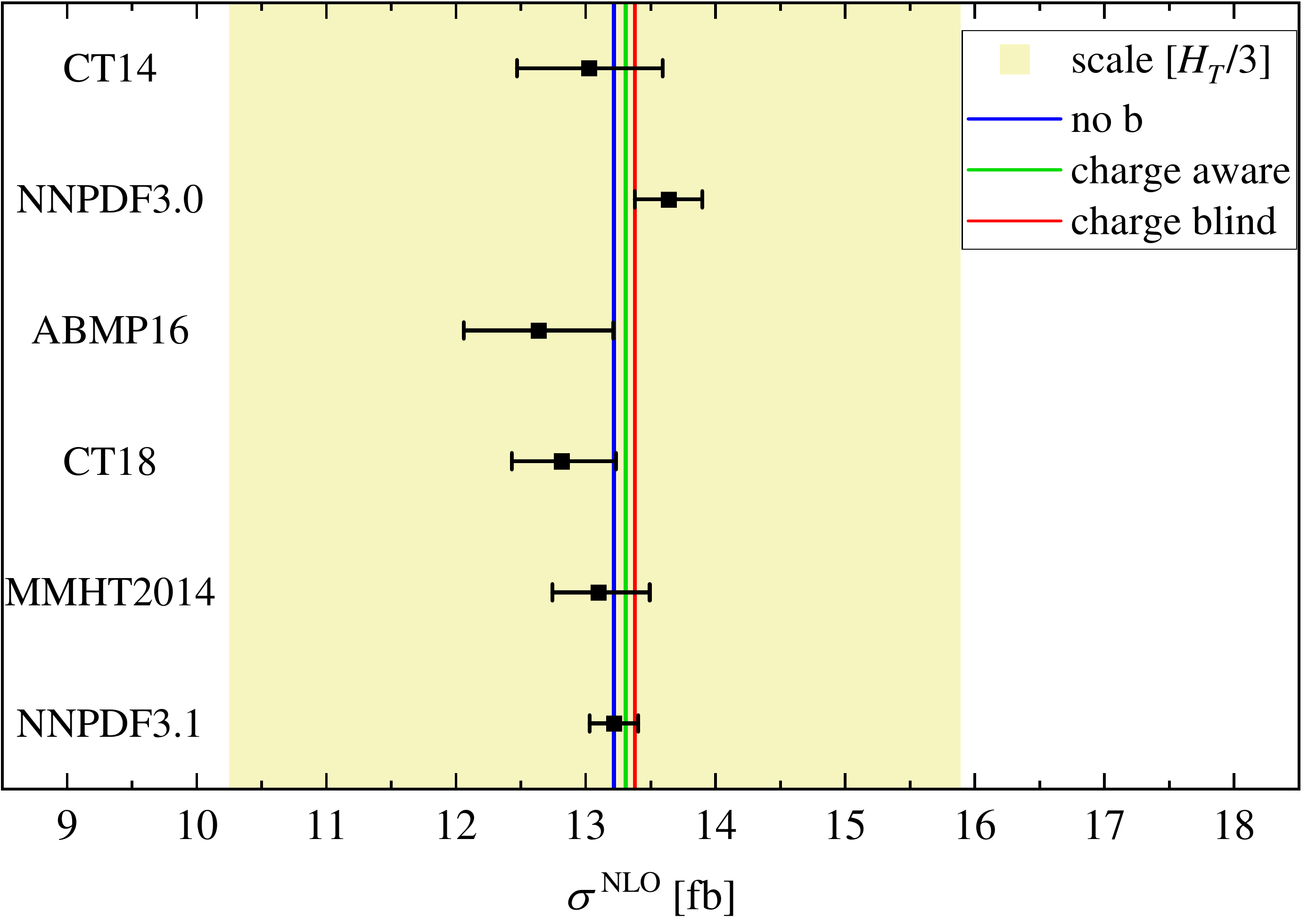}
\caption{\textit{NLO integrated fiducial cross sections for $pp\rightarrow e^+ \nu_e \mu^- \bar{\nu}_\mu b \bar{b} b \bar{b} + X$ at $\sqrt{s}=13$ TeV LHC. We report results with and without $b$-quark initiated subprocesses. We also report predictions obtained with several PDF sets. Figure taken from \cite{Bevilacqua:2021cit}.}}
\label{fig:totxsunc}
\end{figure}

NLO QCD corrections are very important for $pp\rightarrow e^+ \nu_e \mu^- \bar{\nu}_\mu b \bar{b} b \bar{b} + X$. Indeed, these corrections are large and the theoretical uncertainty drops when going from LO to NLO. Our predictions are in very good agreement with the ATLAS experimental results for the integrated fiducial cross section. We also investigated the contribution of the $b$-quark initiated subprocesses, that turned out to be negligible. To conclude, Figure~\ref{fig:totxsunc} reports the various sources of uncertainty for this process. Predictions including or not $b$-quark initiated subprocesses are provided as well as predictions obtained with different PDF sets \cite{Harland-Lang:2014zoa, Dulat:2015mca, NNPDF:2017mvq, Alekhin:2018pai, Hou:2019efy} with their internal uncertainty. All of these uncertainties are negligible compared to the one coming from the scale dependence.

\Acknowledgements
The research of M.L. was supported by the DFG under grant 400140256 - GRK 2497: \textit{The physics
of the heaviest particles at the Large Hardon Collider}.


\providecommand{\href}[2]{#2}\begingroup\raggedright\endgroup

\end{document}